\documentclass[sigconf]{acmart}

\usepackage{subcaption}
\usepackage{algorithm}
\usepackage{graphicx}
\usepackage{algorithmic}
\usepackage{enumitem}
\usepackage{makecell}
\usepackage{multirow}
\usepackage{setspace}
\newcommand{\ri}{\text{(i)~}}
\newcommand{\rii}{\text{(ii)~}}
\newcommand{\riii}{\text{(iii)~}}
\newcommand{\riv}{\text{(iv)~}}
\newcommand{\model}{{{\tt BNCL}}}

\AtBeginDocument{%
  \providecommand\BibTeX{{%
    \normalfont B\kern-0.5em{\scshape i\kern-0.25em b}\kern-0.8em\TeX}}}

\copyrightyear{2023}
\acmYear{2023}
\setcopyright{acmlicensed}\acmConference[CIKM '23]{Proceedings of the 32nd
ACM International Conference on Information and Knowledge
Management}{October 21--25, 2023}{Birmingham, United Kingdom}
\acmBooktitle{Proceedings of the 32nd ACM International Conference on
Information and Knowledge Management (CIKM '23), October 21--25, 2023,
Birmingham, United Kingdom}
\acmPrice{15.00}
\acmDOI{10.1145/3583780.3615039}
\acmISBN{979-8-4007-0124-5/23/10}

\begin{document}



\title{Robust Basket Recommendation via Noise-tolerated Graph Contrastive Learning}

\author{Xinrui He$^*$}
\thanks{$^*$Both authors contributed equally to the paper.}
\affiliation{\institution{University of Illinois at Urbana Champaign}
\country{USA}}
\email{xhe33@illinois.edu}

\author{Tianxin Wei$^*$}
\affiliation{\institution{University of Illinois at Urbana Champaign}
\country{USA}}
\email{twei10@illinois.edu}

\author{Jingrui He}
\affiliation{\institution{University of Illinois at Urbana Champaign}
\country{USA}}
\email{jingrui@illinois.edu}


\renewcommand{\shortauthors}{Xinrui He, Tianxin Wei, \& Jingrui He}

\begin{abstract}

The growth of e-commerce has seen a surge in popularity of platforms like Amazon, eBay, and Taobao. This has given rise to a unique shopping behavior involving baskets – sets of items purchased together. As a less studied interaction mode in the community, the question of how should shopping basket complement personalized recommendation systems remains under-explored. While previous attempts focused on jointly modeling user purchases and baskets, the distinct semantic nature of these elements can introduce noise when directly integrated. This noise negatively impacts the model's performance, further exacerbated by significant noise (e.g., a user is misled to click an item or recognizes it as uninteresting after consuming it) within both user and basket behaviors.
In order to cope with the above difficulties, we propose a novel \textbf{B}asket recommendation framework via \textbf{N}oise-tolerated \textbf{C}ontrastive \textbf{L}earning, named \model, to handle the noise existing in the cross-behavior integration and within-behavior modeling. First, we represent the basket-item interactions as the hypergraph to model the complex basket behavior, where all items appearing in the same basket are treated as a single hyperedge. Second, cross-behavior contrastive learning is designed to suppress the noise during the fusion of diverse behaviors. Next, to further inhibit the within-behavior noise of the user and basket interactions, we propose to exploit invariant properties of the recommenders w.r.t augmentations through within-behavior contrastive learning. A novel consistency-aware augmentation approach is further designed to better identify the noisy interactions with the consideration of the above two types of interactions. Our framework \model~offers a generic training paradigm that is applicable to different backbones. Extensive experiments on three shopping transaction datasets verify the effectiveness of our proposed method. Our code is available at \url{https://github.com/Xinrui17/BNCL}.

\end{abstract}




\maketitle

\begin{figure}[t]
\centering
  \includegraphics[width=0.82\columnwidth]{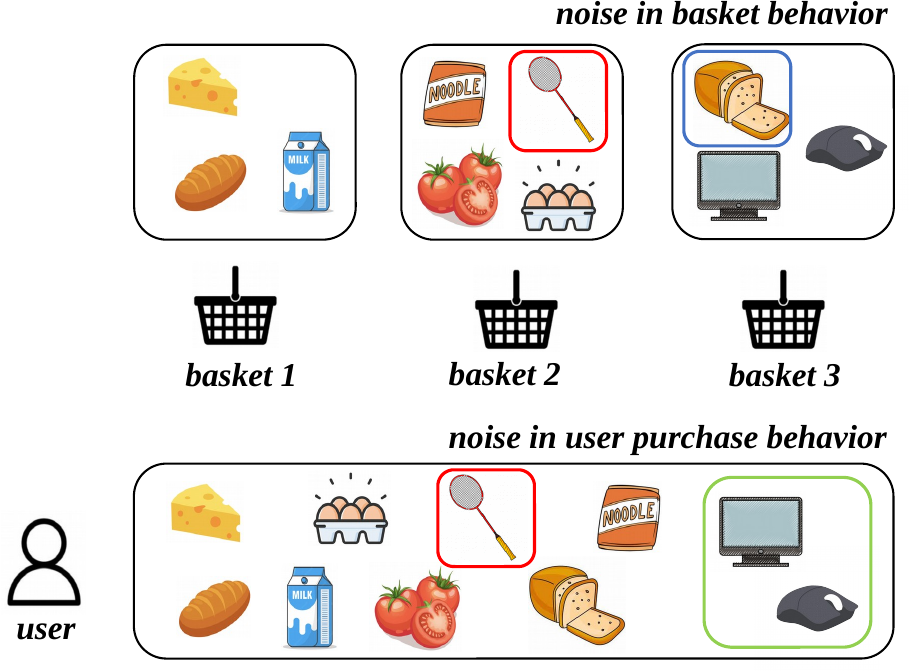}
  \caption{An example of the noise in the basket recommendation: in terms of content relevance in the basket, the food in basket 3 is the noisy item; from the perspective of the users' shopping preferences, it is the electronics in basket 3 that might be treated as the noise. From both two perspectives, the racket shows up as a consistent noise.
  } 
\label{fig:intro}

\end{figure}

\section{Introduction}

Recommender systems have become a powerful tool that greatly enhances shopping experiences on online platforms ever since their inception \cite{balabanovic1997fab,adomavicius2005toward}. 
In practice, customers often purchase multiple items at the same time, and the co-occurrence relationships could provide rich information in item properties mining. Basket recommendation \cite{wan2018representing} is to predict a set of relevant items that a customer will be interested in by analyzing the composition of the historical interactions and the current shopping baskets if given \footnote{A customer may go to this supermarket to buy products many times during a period. Then this customer has multiple shopping baskets}, which can be better used for product arrangement, procurement, promotion, and marketing \cite{gatzioura2014case} to improve customer experience and generate business value. 

There are an increasing number of works \cite{guidotti2018personalized,le2019correlation,benson2016modeling} trying to explore the customer's order history to capture the users' shopping preference and item semantics, with the aim of improving the recommendation quality and boosting online service. However, not all of the order details in the transaction data are essential and relevant to determining the user's next action. There are usually some user-item interactions as well as basket-item interactions appearing as noise due to the diversity of the basket contents and the users' mismatched behaviors. In practice, the noisy interaction occurs in the shopping transaction data for the reason that the users' shopping behavior is somewhat random and fragmented, meaning that users sometimes buy items against their shopping habits or there are some items in the basket that are not related to any others. The existence of such noisy interactions is verified \cite{qin2021world,wu2021self,yang2022knowledge} to hinder the understanding of the user's behavior patterns, which will harm the recommender system training and thus hinder its practical deployment.

Therefore, it is necessary to denoise in basket recommendation (BR) to extract effective information to enhance recommendation performance. However, most existing BR methods \cite{bai2018attribute,wang2015learning,le2018modeling} mainly tend to jointly model purchases and baskets which will introduce noise to the learned representations due to the heterogeneity of user purchases and basket behaviors. To suppress the noise in recommender systems, current works can be classified as follows: \ri The first line of work \cite{wu2021self,ma2022crosscbr,yang2022knowledge} focuses on improving model robustness \cite{sun2023enhance} against user interaction noises.; \rii Another emerging series of papers \cite{qin2021world,le2017basket} put their attention on inhibiting the negative effect of noisy basket behaviors. Both the two types of works only consider the noise within one behavioral pattern, while the joint noise handling of the above two behaviors remains unexplored. What's more, the semantic mismatch of user purchases and basket behaviors will introduce additional noise during the behavioral fusion process. Consequently, it's essential to design the denoise method from a global view that considers the comprehensive information from both two behaviors. To illustrate the noise, we also give an example in Figure \ref{fig:intro}, which presents the consuming history of a user from the view of user purchase behavior (below) and the view of basket behavior (above). We observe that, from the basket behavior view, the bread in the basket $3$ can be regarded the outlier with high probability. However, if we take a look at the user purchase behavior, the bread could not be treated as an anomaly while the computer and mouse are more likely to noise for the user. Moreover, if we look at the racket, which is both unimportant in the two views, there is a high probability for it to be a consistent noise. From the example, we're inspired that it's crucial to design the denoise approach that considers information comprehensively from both within and across the two behavior views.

In this paper, we propose a comprehensive within-basket recommendation framework \model~ via noise-tolerated contrastive learning to handle the noise existing in the cross-behavior integration and within-behavior modeling. To be more specific, first, we adopt the typical user-item bipartite graph to model the user-item interactions from the user purchase behavior, and the basket-item interactions for the basket behavior are represented as the hypergraph to model the complex basket behavior, where all the items appearing in the same basket are treated as a single hyperedge. Secondly, we propose cross-behavior contrastive learning to fuse the representations learned from the basket behavior into recommender systems, which aims at suppressing the noise introduced by the fusion of diverse behaviors. Then, to handle the within-behavior noise of the user and basket interactions respectively, we propose to exploit invariant properties of the recommenders w.r.t augmentations through within-behavior contrastive learning. During the process, a novel consistency-aware augmentation method is proposed to better identify the noisy interactions with the comprehensive information of both the two types of behaviors. 
To optimize the model, we leverage a multi-task training strategy to jointly optimize the classic recommendation task and the self-supervised contrastive denoising task. In summary, the contributions of this paper could be summarized as follows:

\begin{itemize}[leftmargin=*]

\item This work formulates the idea of integrating the basket behaviors into user-item interaction modeling with a light hypergraph message passing schema as well as a joint self-supervised learning paradigm.

\item We systematically illustrate the noise issues in the within-basket recommendation problem and propose a general basket recommendation framework \model~ to improve robustness against the within-behavior and cross-behavior noise issues via noise-tolerated contrastive learning.

\item Extensive experimental results over three shopping transaction datasets show that our proposed method outperforms state-of-the-art baselines in terms of various ranking metrics.

\end{itemize}
The rest of the paper is organized as follows. We show the preliminary definition in Section \ref{section:preliminary} and introduce the proposed \model~ in Section \ref{section:method}. Then we present the experimental results in Section \ref{section:experiments}. Section \ref{section:related_work} briefly discusses the existing work. In the end, we conclude the paper in Section \ref{section:conclusion}.

\begin{figure}[t]
\centering
  \includegraphics[width=1\columnwidth]{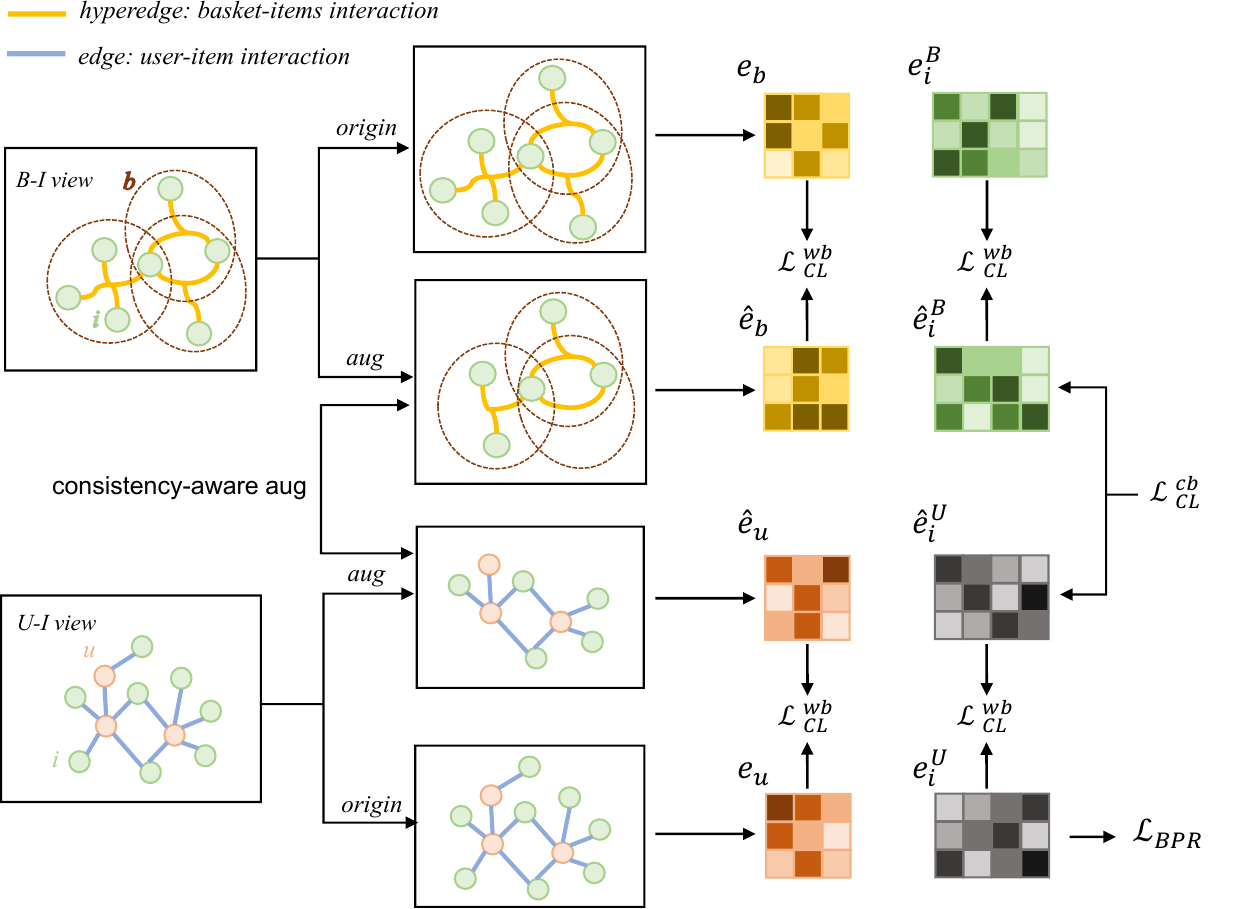}
  \caption{The overall framework of \model. \ri~In the view of basket behavior, the yellow hyperedges denote the baskets and the nodes denote the items. In the user purchase behavior, the blue edges represent the user-item interactions. \rii~We adopt within-basket contrastive learning between the augmented graph and the original graph for this two behavior respectively. \riii~The augmented views of the user purchase behavior and the basket behavior are obtained by the proposed consistency-aware augmentation. \riv~Cross-behavior contrastive learning helps to better fuse the item properties learned from two behaviors.}
\label{fig:framework}

\end{figure}

\section{Preliminary}
\label{section:preliminary}

\subsection{Within-basket Recommendation Setting}
As a common practice, we use $U=\{u_1, u_2,\dots, u_{|U|}\}$ to represent all users and $I=\{i_1, i_2, \dots, i_{|I|}\}$ to represent all items where $|U|$ and $|I|$ denote the number of the users and the items respectively. We consider the basket as a set of items that the user ordered in one transaction. Therefore, we will obtain an interaction basket sequence according to the transaction record of the user $u$, which is denoted as $B^{u}=\left(b_{1}^{u}, b_{2}^{u}, \ldots, b_{\left|B^{u}\right|}^{u}\right)$ where $\left|B^{u}\right|$ is the number of the baskets that user $u$ has purchased and $b_{j}^{u} \subseteq I$ represents the $j^{th}$ basket purchased by user $u$. Within-basket recommendation task aims to recommend the most possible item list to be added to a partially given basket $b_{p}^{u}$ associated with a user u.
\subsection{User-item Interaction View Learning}
In the raw transaction data, each user and item is assigned a unique ID respectively. We use a $d$ dimension embedding to represent each user and item where $\mathrm{e}_{u} \in \mathbb{R}^{d}$ is the embedding of the user, $\mathrm{e}_{i} \in \mathbb{R}^{d}$ is the embedding of the item. To capture the users' purchase behavior, we adopt the idea of the user-item bipartite graph to model the user-item interactions, in which the user and the item are treated as nodes. If the user $u$ has bought the item $i$, there will be an edge connecting these two nodes on the user-item interaction graph. 

We adopted the graph convolutional network LightGCN \cite{he2020lightgcn} to perform the message passing on the user interaction graph. 

Following LightGCN, the $k$-th layer information propagation on the user-item bipartite graph could be described as follow:
\begin{equation}
    \left\{\begin{array}{l}\mathbf{e}_{u}^{(k)}=\sum_{i \in \mathcal{N}_{u}} \frac{1}{\sqrt{\left|\mathcal{N}_{u}\right|} \sqrt{\left|\mathcal{N}_{i}\right|}} \mathbf{e}_{i}^{U(k-1)} \\ \mathbf{e}_{i}^{U(k)}=\sum_{u \in \mathcal{N}_{i}} \frac{1}{\sqrt{\left|\mathcal{N}_{i}\right|} \sqrt{\left|\mathcal{N}_{u}\right|}} \mathbf{e}_{u}^{(k-1)}\end{array}\right.
\end{equation}
where $\mathrm{e}_{u}^{(k)}$, $\mathrm{e}_{i}^{U(k)} \in \mathbb{R}^{d}$ are the embedding of the user $u$ and the item $i$ at $k$-th layer respectively. We randomly initialize the user and item embedding $\mathrm{e}_{u}^{(0)}$, $\mathrm{e}_{i}^{U(0)}$ at the very beginning. 

To take the aggregating information at different depths into account, the final representations $\mathrm{e}_{u}$ and $\mathrm{e}_{i}^{U}$ of the user and the item are obtained as the mean of the output embedding of different layers:
\begin{equation}
    \mathrm{e}_{u}=\frac{1}{K}\sum_{k=0}^{K} \mathrm{e}_{u}^{k}
\end{equation}
\begin{equation}
  \mathrm{e}_{i}^U=\frac{1}{K}\sum_{k=0}^{K} \mathrm{e}_{i}^{U(k)}
\end{equation}
where $K$ is the number of the layers, $\mathrm{e}_{u}$ and  $\mathrm{e}_{i}$ are the final representations of the user and the item.

\subsection{Contrastive learning}
Intuitively, the different views of the same sample are likely to be clustered in the embedding space. Contrastive learning (CL) aims at learning good data representations by reducing the distance of the defined positive pairs while pushing away the representations of the negative pairs in the embedding space. The main process of CL \cite{wang2020understanding} is to first construct the diverse views of the raw data through an augmentation set $\mathcal{A}$. Given a data point $\boldsymbol{x}$, the augmentation results of $\boldsymbol{x}$ are denoted as $\mathcal{A}(\boldsymbol{x}$), which are treated as the positive pairs, otherwise, the negative pairs. CL tries to learn an encoder $f$ such that the positive pairs are well aligned and the negative ones have been pushed apart. We denote the $\boldsymbol{z(x)}$ as the representation of sample $\boldsymbol{x}$ in the embedding space. To achieve this goal, a class of methods employed InfoNCE \cite{oord2018representation,caron2020unsupervised,chen2021exploring} as the contrastive loss function, formulated as:
\begin{equation}
\mathcal{L}_{\text {InfoNCE }}=\sum_{x \in \mathcal{X}, x^\prime \in \mathcal{A}(x)}-\log \frac{\exp \left(\operatorname{sim}\left(x, x^\prime\right) / \tau\right)}{\sum_{x_{k} \in \mathcal{X}} \exp \left(\operatorname{sim}\left(x, x_{k}\right) / \tau\right)}
\end{equation}
where $\mathcal{X}$ is the set of data samples in a mini-batch, $x^\prime$ is the augmented view of a random data point $x$, $\operatorname{sim}(\cdot, \cdot)$ is a similarity function and $\tau$ is the temperature coefficient used to control the uniformity of the representation in the embedding space.
In this case, the model is more likely to learn the more invariant and essential properties of the raw data by mapping the positive pairs into the nearby space. 

To adapt to the downstream task, it is natural to adopt the contrastive objective as a ladder combined with supervised signals to form a multitask learning objective.

\section{Method}
\label{section:method}
In this section, we introduce the proposed \model. The overall framework is shown in Figure \ref{fig:framework}. First, we model the user-item and basket-item interactions from the user purchase behavior and the basket behavior through a user-item interaction graph and a complementary basket hypergraph respectively. Then, we propose cross-behavior contrastive learning to eliminate the noise introduced in the fusion of the representations learned from purchase and basket behavior. Finally, within-behavior contrastive learning is employed in the user’s purchase
behavior and the basket behavior respectively for denoising these two behaviors. In addition, a consistency-aware augmentation is designed in the
within-behavior contrastive learning to help identify the true
noisy interactions.
\subsection{Complementary View Learning}

In the basket recommendation task, the orders are usually unique to the users, which means it is not likely that two customers share exactly the same basket. Compared with user-basket interactions, what matters most is the interactions between users and items, which contains the cross-basket information, as well as the interrelated information among items within a basket. Thus, we propose to learn the representations from both the user purchase behavior and the basket behavior.

For the user purchase behavior, which contains the user-item interactions, we employ the user-item interaction graph and perform LightGCN message passing on it to learn the representations of the user and item, denoted as $\mathrm{e}_{u}$ and $\mathrm{e}_{i}$. From the user purchase behavior, we encode the user's personalized information and shopping preferences to the user and item embedding.

Different from the general recommendation task, the basket recommendation scenario is provided with the shopping basket records which contain valuable correlation information about the items. Specifically, the compositions in a basket often imply the complementarity, similarity, or substitution relationships among the item. The basket behavior is encoded in the basket-item interactions, which serves as a supplementary part to extract effective item properties for recommendation.
We use a basket hypergraph $\mathcal{G}_{hyper}=(V, E)$ to model the basket behavior for the following reasons. \ri~In the hypergraph, a hyperedge can connect two or more vertices \cite{zhu2020hgcn}, which can well model the diverse relationship between a basket and multiple items. \rii~The hypergraph has the ability to aggregate high-order information through the message passing among hypernodes and hyperedges, which is essential to capturing the complex relations between various baskets and items.
 Here $V$ is the set of vertices and each vertice represents an item in the item set. We use $\epsilon$ to denote the basket-item interactions, which means that a hyperedge $\epsilon \in E$ denotes a basket $b$ and the vertices it connects represent all the items belonging to this basket. The basket hypergraph contains $|I|$ vertices and $M$ hyperedges.  The relationship between vertices and hyperedges can be described by an incidence matrix $H \in \mathbb{R}^{|I| \times M}$ defined as follows:
 \begin{equation}
\mathbf{H}(v, \epsilon)=\left\{\begin{array}{ll}1, & \text { if } \epsilon \in E \\ 0, & \text { otherwise }\end{array}\right.
 \end{equation}
Following the spectral hypergraph convolution proposed in \cite{feng2019hypergraph}, we design light message passing on the constructed hypergraph to effectively perform the information aggregation with hyperedges as the mediators. The representation of vertices ${E}_{I}^{B(k)}$ at $k^{th}$ layer is obtained as :
\begin{equation}
    \mathbf{E}_{I}^{B(k)}=\mathbf{D}^{-1 / 2} \mathbf{H B}^{-1} \mathbf{H}^{T} \mathbf{D}^{-1 / 2} \mathbf{E}_{I}^{B(k-1)}
\end{equation}
where $D \in \mathbb{R}^{|I| \times |I|}$ and $B\in \mathbb{R}^{M \times M}$ are the diagonal degree matrices of the vertices and the hyperedges, where each entry denotes the degree value of corresponding items/hyperedges. ${E}_{I}^{B(k)} \in \mathbb{R}^{|I| \times d^k}$ is the item embedding matrix at the $k^{th}$ layer on the hypergraph and $d^k$ is the hidden size of $k^{th}$ layer. We use ${e}_{i}^{B(k)}$ to denote the representation of item $i$ at the $k^{th}$ layer, which is equal to the $i^{th}$ row of ${E}_{I}^{B(k)}$.

We use the mean of the representations at different layers as the final representation of the items learned from the basket behavior:
\begin{equation}
    \mathrm{e}_{i}^{B}=\frac{1}{K}\sum_{k=0}^{K} \mathrm{e}_{i}^{B(k)}
\end{equation}
Defining $\mathcal{N}_b=\{i\mid \mathbf{H}(i,b)=1\}$ as the items incorporated in the basket $b$, we can obtain the final basket embedding from the items:
\begin{equation}
    \mathrm{e}_{b}=\frac{1}{|\mathcal{N}_b|}\sum_{i\in \mathcal{N}_b} \mathrm{e}_{i}^{B}
\end{equation}

\subsection{Cross-behavior Contrastive Learning}
Item embedding $\mathrm{e}_{i}^{U}$ and $\mathrm{e}_{i}^{B}$, learned from the user purchase behavior and the basket behavior, capture inherent semantic properties of different views. However, loosely integrating them together may lead to suboptimal results due to the inconsistent feature space and view aggregation schema. To accomplish the fusion of diverse behaviors, in this section, we propose cross-behavior contrastive learning to suppress the noise during the fusion of the user purchase behavior and the basket behavior.

Motivated by the core idea of CL, we introduce cross-behavior contrastive learning to align the representation of the same item generated from different behaviors, which could unify the item embedding into the same feature space and further explore the intrinsic semantic properties of the item.

The representation of the same item should share the same intrinsic semantic properties. Thus, we define the positive pairs as the embedding of the same item learned from the augmentation views of the user purchase behavior and the basket behavior, denoted as $\hat{\mathrm{e}}_{i}^{U}$ and $\hat{\mathrm{e}}_{i}^{B}$. Then we treat the representations of different item embedding learned from different behaviors as negative pairs. Employing the InfoNCE \cite{oord2018representation}, the objective of cross-behavior contrastive learning could be formulated as:
\begin{equation}
    \mathcal{L}_{CL}^{cb}=\sum_{i \in \mathcal{I}}-\log \frac{\exp \left(\operatorname{sim}\left(\hat{\mathrm{e}}_{i}^{U}, \hat{\mathrm{e}}_{i}^{B}\right) / \tau\right)}{\sum_{j \in I, i \neq j} \exp \left(\operatorname{sim}\left(\hat{\mathrm{e}}_{i}^{U}, \hat{\mathrm{e}}_{j}^{B}\right) / \tau\right)}
\end{equation}
where $\operatorname{sim}(\cdot, \cdot)$ is cosine similarity function, $\tau$ is the temperature of the contrastive learning.

\subsection{Within-behavior Contrastive Learning}
In the above section, we model the user purchase behavior and the basket behavior through a user-item interaction graph and a basket hypergraph and learn the representation from these two graphs. However, in real-world shopping transaction records, there are always some items that are irrelevant to the user preference or to the intents of the basket which appear as noise. The existence of noise will affect the learned user shopping preferences and deteriorate the quality of the representation.

To tackle this problem, we propose within-behavior contrastive learning on the user purchase behavior and the basket behavior respectively to achieve the goal of denoising the user-item interactions and the basket-item interactions. To be more specific, we generate the augmented view of the user-item interaction graph and the basket hypergraph. Then, the message passing is performed both on the original graph and the augmented graph. We denote the representation of the user and item on the user-item interaction graph as $\mathrm{e}_{u}$ and $\mathrm{e}_{i}^{U}$, the representation of the basket and item on the basket hypergraph as $\mathrm{e}_{b}$ and $\mathrm{e}_{i}^{B}$. Also, we have the representation of the user and item $\hat{\mathrm{e}}_{u}$, $\hat{\mathrm{e}}_{i}^{U} $ on the augmented view of the user-item interaction graph, and the representation of the basket and item $\hat{\mathrm{e}}_{b}$, $\hat{\mathrm{e}}_{i}^{B}$ on the augmented view of the basket hypergraph.
The objective of the user representation learning in within-behavior contrastive learning is defined as:
\begin{equation}
    \mathcal{L}_{CL}^{U}=\sum_{u \in \mathcal{U}}-\log \frac{\exp \left(\operatorname{sim}\left(\mathrm{e}_{u}, \hat{\mathrm{e}}_{u}\right) / \tau\right)}{\sum_{v \in U, u \neq v} \exp \left(\operatorname{sim}\left(\mathrm{e}_{u}, \hat{\mathrm{e}}_{v}\right) / \tau\right)}
\end{equation}

We adopt the same contrastive objectives for item embedding of the original view and the augmented view on the user-item interaction graph, and the basket embedding as well as the item embedding of the original view and the augmented view on the basket hypergraph. Four contrastive terms are obtained as $\mathcal{L}_{CL}^{U}$, $\mathcal{L}_{CL}^{I_u}$, $\mathcal{L}_{CL}^{B}$, $\mathcal{L}_{CL}^{I_b}$.
The final contrastive learning loss for within-behavior denoising is the sum of these four terms:
\begin{equation}
    \mathcal{L}_{CL}^{wb}= \mathcal{L}_{CL}^{U}+\mathcal{L}_{CL}^{I_u}+\mathcal{L}_{CL}^{B}+ \mathcal{L}_{CL}^{I_b}
\end{equation}

\subsection{Consistency-aware Augmentation}
\label{subsection:consistency aug}
Up to now, we have investigated denoising techniques applicable for each behavior respectively, as well as during the fusion of heterogeneous behaviors.
However, developing an effective augmentation strategy for contrastive learning that precisely identifies the noisy interactions, considering these two types of behaviors, remains an unresolved challenge.

Commonly used data augmentation methods for graph structures, such as stochastic edge and node removal, introduce substantial randomness in altering the graph's structure and identifying noise. For example, when the edges removed are connected to important nodes, some fundamental relationships may be lost and the underlying structure of the graph may be destroyed as a result. Moreover, previous augmentation techniques that rely on a single view can be unreliable in the context of basket recommendation. An item can be considered noise from one perspective, while it may carry meaningful information when viewed from another perspective, as depicted in Figure \ref{fig:intro}. To mitigate the risk of losing important relationships, it is crucial to develop an augmentation strategy that integrates crucial multi-view information and removes interactions that consistently exhibit noise.

Aiming to identify the noise, we propose a consistency-aware augmentation approach for cross-behavior and within-behavior contrastive learning. To be more specific, we comment that a user-item interaction or a basket-item interaction tends to be noisy when the item is considered potentially noisy based on both the user purchase behavior and the basket behavior. Inspired by the definition of the node centrality in the graph \cite{10.1093/acprof:oso/9780199206650.001.0001}, we define the interaction importance $s_{u-i }^{\epsilon}$ and $s_{b-i }^{\epsilon}$ upon two graphs as:

\begin{equation}
    s_{u-i }^{\epsilon}=\log\left(\delta_{d}(u)+\delta_{d}(i) + \delta_{d\_hyper}(i) \right) 
 \end{equation}
\begin{equation}
    s_{b-i }^{\epsilon}=\log\left(\delta_{d\_hyper}(b)+\delta_{d\_hyper}(i) + \delta_{d}(i) \right) 
\end{equation}
where $ \delta_{d}(u)$ and $ \delta_{d}(i)$ are the degree of the user and item node on the user-item interaction graph. On the basket hypergraph, we focus on both the degree of the hyperedge $ \delta_{d\_hyper}(b)$ which is defined as the number of items contained on the basket hyperedge, and the degree of the item vertex $ \delta_{d\_hyper}(i)$ which is defined as the number of hyperedges connecting to this item vertex. 

According to the expression of edge importance, the importance of an edge is defined from the perspective of the user purchase behavior as well as the basket behavior. We comment that a user-item interaction or a basket-item interaction is more important when it has greater edge importance on the corresponding graph. To generate more meaningful augmentation, we are inclined to drop the less important interactions.

To calculate the probability of dropping an edge $\epsilon$, for example, on the user-item interaction graph, we use normalization to transform the edge importance $s_{u-i}^{\epsilon}$ of edge $\epsilon$ into the probability:
\begin{equation}
    p_{\epsilon}=\frac{s_{\max }-s_{u-i}^{\epsilon}}{s_{\max }-s_{\min}} \cdot p
\end{equation}
where $p$ is the overall edge drop probability, $s_{\max }$ and $s_{\min }$ is the max and min value of $s_{u-i}$. Following the same formula, the probability of dropping a basket-item interaction $\epsilon$ on the basket hypergraph can be obtained as well. According to the probability $p_{\epsilon}$ of dropping the edge $\epsilon$, we could obtain the augmentation view of the original graph $\mathcal{G}$ where the probability that the edge $\epsilon$ belongs to the augmentation view $\hat{\mathcal{G}}$ is $1-p_{\epsilon}$.

With consistency-aware augmentation, the positive pairs generated from two behavior views tend to share more common intrinsic properties for better alignment in cross-behavior contrastive learning. While it can generate the more meaningful augmentation view for within-behavior contrastive learning on the user-item interaction graph and basket hypergraph respectively.

\subsection{Prediction and Optimization}
To perform recommendations, the ranking score of each user-item pair considers both the user information and current basket information:

\begin{equation}
    \hat{y}_(u,b,i)=(1-r)\cdot\mathbf{e}_{u}^{ \top} \mathbf{e}_{i}+r\cdot \mathbf{e}_{b}^{ \top}\mathbf{e}_{i}
    \label{score2}
\end{equation}

where $\mathbf{e}_{i}=\mathbf{e}_{i}^{U}+\mathbf{e}_{i}^{B}$ denotes the fused item embedding performed with contrastive denoising, $r$ is a hyperparameter to balance the capacity of the user and the basket in the recommendation.

We use the BPR loss \cite{rendle2012bpr} as the main recommendation loss. We sample a positive item $i$ and a negative item $j$ for the user $u$, where the positive item is selected within the current basket and the negative item is sampled from items without being purchased. The main loss is:
\begin{equation}
    L_{\text {main }}=-\sum_{\left(u, i, j\right)} \log \sigma\left(\hat{\mathbf{y}}_{\left(u,b, i\right)}-\hat{\mathbf{y}}_{\left(u, b,j\right)}\right)+\lambda\|\Theta\|_{2}^{2}
\end{equation}
where $\Theta$ is the model parameters and $\lambda$ is a positive constant. To improve recommendation with the self-supervised denoising tasks, we leverage a multi-task training strategy to jointly optimize the classic recommendation task, the cross-behavior contrastive task and the within-behavior contrastive task. In this way, we have the final multi-task loss as:
\begin{equation}
    \mathcal{L}= \mathcal{L}_{main}+\alpha_{1} \mathcal{L}_{CL}^{cb}+\alpha_{2} \mathcal{L}_{CL}^{wb}
\end{equation}
where $\alpha_1$ and $\alpha_2$ are hyperparameters to control the linear weight.

\renewcommand\arraystretch{1.3}
\begin{table}[t]
\setlength{\abovecaptionskip}{0.2cm}
\caption{Dataset Statistics.}
\label{table:statistic}
\scalebox{0.9}{
\begin{tabular}{ccccc}
\hline 
              & \#Users         & \#Items        & \makecell[c]{Average \\ Basket Size}         & \makecell[c]{Average \\ \#Baskets per User}          \\ \hline
Instacart         & 22168          & 40044          & 37.00         & 2.96          \\ 
Tafeng         & 7119         & 11916         & 15.99        & 2.20           \\
Valuedshoppers         & 9532          & 7860          & 19.94          & 17.66         \\
\hline
\end{tabular}}

\end{table}

  \begin{table*}[ht]

\small
\caption{Experimental results on the three real-world datasets through different methods with \% omitted. The best
results are highlighted in boldface. Underlined values indicate the second best.}
\centering
\resizebox{1\textwidth}{!}{
\begin{tabular}{lc|c|cc|cc|ccc|cc|c}
\toprule
 \multirow{2}{*}{Data}&\multirow{2}{*}{Metric}&\multicolumn{1}{c|}{Simple methods} &\multicolumn{2}{c|}{MF-based methods}&\multicolumn{2}{c|}{NBR methods}&\multicolumn{3}{c|}{GNN-based methods}&\multicolumn{2}{c|}{Denoising methods}&\multirow{2}{*}{BNCL}\\
  &&PersonPop-k&BPRMF &Triple2Vec &DREAM &TIFU-KNN &LightGCN &Basconv &MITGNN &SGL &CLEA  \\

\midrule
\multirow{6}{*}{Instacart}
&Recall@40 &1.959&12.005& 9.285 &9.430 &13.567  &13.792 &12.569 & 11.461& 15.742 &\underline{15.919} & \textbf{17.156} \\
&Recall@60 &2.309&15.612& 11.711 & 11.105& 20.407 &17.832 &17.144 & 15.426& 19.350 &\underline{20.701}& \textbf{21.099} \\
&HR@40 &13.377&56.928 & 46.405 & 44.542& 54.174&61.355 &57.725 & 64.705&63.423 &\underline{65.334} &\textbf{67.169}\\
&HR@60 &15.433&66.048 &53.930 &49.553 & 74.197&70.313 &68.690 & 64.985&70.286 & \underline{74.356} &\textbf{74.674}\\
&NDCG@40 &4.127&14.451 & 14.359 &12.061 & 5.794&18.463 &15.157 & 14.526&21.492&\underline{19.390} & \textbf{22.742} \\
&NDCG@60  &4.541&19.876 & 16.474 &13.440 & 9.263&21.504 &18.581 & 17.630&24.162&\underline{22.760} & \textbf{25.615} \\
\midrule
\multirow{6}{*}{TaFeng}
&Recall@40 &0.263&2.075&0.082 &0.381 &2.459&3.178 &2.799 &2.745 &3.283 & \underline{3.382} & \textbf{3.488} \\
&Recall@60 &0.468&2.648&0.188 & 0.513&2.836&3.781 &3.437&3.467 &3.893 &\underline{4.062} &\textbf{4.175} \\
&HR@40 &1.063&6.634 & 0.300 & 1.634& 8.586&10.200 &9.052&8.867 & 10.583 &\underline{10.755} &\textbf{11.845}\\
&HR@60 &1.905&8.369 & 0.612 &2.177 & 9.470&11.948 &10.966& 10.934&12.356 &\underline{12.607} &\textbf{12.835}\\
&NDCG@40 &0.234&2.051 &0.078& 0.321& 0.870&3.898& 2.879&2.497  &4.000 &\underline{4.054} & \textbf{4.356}\\
&NDCG@60  &0.389&2.378 &0.134& 0.413&0.978 &4.244& 3.257&2.902 &4.351 &\underline{4.428} &\textbf{4.628}\\
\midrule
\multirow{6}{*}{ValuedShopper}
&Recall@40 &1.063&1.963& 1.655 &1.638 & \underline{6.567}&5.844 &5.476 &5.903 &5.995 & 5.384 &\textbf{6.659}\\
&Recall@60 &1.486&2.873& 2.856 &1.979 &\underline{8.206} &7.780 &7.281& 7.619&8.045 &7.505 &\textbf{8.845}\\
&HR@40 &3.910&6.359& 5.417&3.161 & 20.219& 19.600 &18.413& \underline{20.275}&20.216 &17.701 &\textbf{22.136}\\
&HR@60 &5.494&9.324 &9.417& 4.336 & \underline{26.521}&25.221 &23.690& 25.234&26.224 & 23.822 &\textbf{28.334}\\
&NDCG@40 &1.092&1.840 & 1.747&1.579 & 2.581&5.489& 5.485& \underline{6.035}&5.659& 4.766 &\textbf{6.364}\\
&NDCG@60  &1.374&2.408 & 2.505& 1.788&3.253 &6.634&6.572&\underline{7.153} &6.882& 6.008 &\textbf{7.529}\\
\bottomrule
\end{tabular}
}
\label{tab:main}

\end{table*}

\section{Experiments}
In this section, we conduct experiments to evaluate the performance of our proposed \model. Our experiments intend to answer the following research questions:
\begin{itemize}[leftmargin=*]
    \item \textbf{RQ1: } How does \model~perform in the within-basket recommendation task compared with the baseline models?
    \item \textbf{RQ2: }How do different components in \model~contribute to the performance?
    \item \textbf{RQ3: }How is the generalization ability of our proposed \model~under different circumstances (e.g., varying length of recommended item list and backbones)?
    \item \textbf{RQ4: }How does the proposed \model~ perform in the presence of noise (the robustness of \model~to the varying ratio of noise added)?
\end{itemize}
\label{section:experiments}

\subsection{Dataset}
We evaluate the within-basket recommendation performance on real-world datasets: Instacart \footnote{https://www.kaggle.com/c/instacart-market-basket-analysis}, Tafeng \footnote{https://www.kaggle.com/chiranjivdas09/ta-feng-grocery-dataset} and Valuedshoppers \footnote{https://www.kaggle.com/c/acquire-valued-shoppers-challenge}.
\begin{itemize}[leftmargin=*]
\item \textbf{Instacart} is a transaction dataset collected from an online shopping grocery. It contains the record of over 3 million grocery orders over time which come from more than 200,000 users.
\item \textbf{Tafeng} contains Chinese grocery store transaction data over four months released by ACM RecSys. It consists of the records of over 13000 users' shopping orders.

\item \textbf{Valuedshoppers} provides almost 350 million purchase histories from over 300,000 shoppers which includes a large set of users' basket-level shopping behaviors. Considering a large number of records, We sampled the transactions for training and prediction.
\end{itemize}
We treat the set of items purchased by a user during a time session as a shopping basket. In order to make the basket informative enough to be useful in the algorithm, we remove baskets containing less than 30 items for Instacart, and less than 10 items for the Valuedshoppers and Tafeng due to the sparsity of the basket-item interactions in these two datasets. The statistics of the final processed datasets are shown in Table \ref{table:statistic}. We split 80\% items of each basket as training data and the remaining 20\% as test data for both of the datasets.

\subsection{Experimental Settings}
\subsubsection{Evaluation metrics} 
We evaluate the performance of models by the Top $K$ recommendation metrics, including the Recall@K, Precision@K, HR@K, and NDCG@K \cite{he2017neural, resnick1997recommender}. We first compute the recommendation score for the given user $u$ with all the items then full ranking is executed to generate the top $K$ most possible items. 
\subsubsection{Baseline}

We consider the following baselines for comparison:

\textbf{Simple method:}
\begin{itemize}[leftmargin=*]
\item \textbf{PersonPop-k}: It is a basic method to return top $k$ items from
the training set in terms of the purchase frequency of a given user.

\end{itemize}

\textbf{MF-based Methods:}
\begin{itemize}[leftmargin=*]
    \item \textbf{BPR-MF} \cite{rendle2012bpr}: It is a method to model user and item interactions. The representation is learned by maximizing the distance between the user and its purchased and unpurchased items.

    \item \textbf{Triple2vec} \cite{wan2018representing}: It learns the user and item representation via the triplets $(item, item, user)$ sampled for a  single shopping basket.
\end{itemize}

\textbf{NBR Methods:}

Additionally, we modify the relevant NBR (Next-Basket Recommendation) methods to suit our within-basket recommendation setting. Specifically, we treat the current partially provided basket as the last basket in the shopping history for next-basket prediction.
\begin{itemize}[leftmargin=*]
    \item \textbf{Dream} \cite{wu2021self}: It leverages recurrent neural networks to model the dynamics of users’ behaviors and the sequential patterns between items. 
    \item \textbf{TIFUKNN} \cite{qin2021world}: It is a nearest neighbor-based model that outperforms deep recurrent neural networks in NBR. It relies on the similarity of the target user with other users and the purchase history of the target user.
\end{itemize}

\textbf{GNN-based Methods:}
\begin{itemize}[leftmargin=*]

    \item \textbf{LightGCN} \cite{he2020lightgcn}: It is a simplified model of NGCF \cite{wang2019neural}. Lightgcn directly uses the normalized summation of neighbors to perform aggregation on the graph, which greatly improves the recommendation performance.

    \item \textbf{BasConv} \cite{liu2020basconv}: It constructs a UBI graph and then designs the heterogeneous aggregators on the graph to realize an informative message passing in representation learning.

    \item \textbf{MITGNN} \cite{liu2020basconv}: It is the recent model focused on the within-basket recommendation task, which retrieves multiple intents across the defined basket graph to learn the representation of users and items.
\end{itemize}

\textbf{Denoising Methods:}
\begin{itemize}[leftmargin=*]
    \item \textbf{SGL} \cite{wu2021self}: It combines the collaborative graph neural network filtering model with contrastive learning for recommendation by perturbing the graph structure through simple data augmentation operations on the graph structure.
    \item \textbf{CLEA} \cite{qin2021world}: CLEA denoises the basket by automatically splitting the basket into positive and negative sub-baskets and using anchor-guided contrastive learning. We adopt the idea of CLEA and adapt the model to the within-basket recommendation task.

\end{itemize}
Note that we omit the comparison with the potential basket recommendation baseline PerNIR \cite{ariannezhad2023personalized} as their objective focuses on predicting the next item for the current basket rather than complementing the entire basket.
\subsubsection{Parameter settings}
For fair comparisons, we adopt the following setting for all methods: the batch size is set to 1024; the embedding size is fixed to 128; all embedding parameters are initialized by Xavier initialization \cite{glorot2010understanding}; the hidden dimension is 64 for all methods. We optimize each baseline method according to the validation set. For LightGCN and SGL, we adopt 3 layers of propagation to achieve their best performance on all datasets. For BasConv model, 2 layers' aggregation is performed to have the best results. For our \model~model, 
We fine-tune the hyperparameter $r$ within the range of $[0, 0.1, 0.2, 0.5]$ and $p$ within the range of $[0.1, 0.3, 0.5, 0.7]$. 
the coefficients for both the cross-behavior and the within-behavior contrastive learning are tuned in the range of $[1\times 10^{-1}, 1\times 10^{-3}, 1\times 10^{-5}]$. Our model converges best when the learning rate is $5\times 10^{-4}$, and the number of propagating layers is 2 for u-i and b-i graph. On Instacart, the coefficients for both cross-behavior and within-behavior contrastive learning are $0.1$; On Tafeng, when the coefficients for cross-behavior and within-behavior contrastive learning are set to $1\times 10^{-2}$ and $1\times 10^{-3}$ respectively, the model reaches the best performance; On Valuedshoppers, we choose the coefficients as $1\times 10^{-4}$ and $1\times 10^{-5}$ respectively for cross-behavior and within-behavior contrastive learning.

\subsection{Results (RQ1) }
In this section, we compare the performance of several state-of-the-art baselines on the within-basket recommendation task on three real-world datasets and the results for $K=40, 60$ are shown in Table \ref{tab:main}. The baselines are arranged according to the different types of models.

    We find that \model~ consistently outperforms other methods, which demonstrates its remarkable ability to extract informative representations by leveraging user-item interactions, basket-item interactions by effectively filtering out irrelevant or noisy information. The reported best-performing models are significant w.r.t. the second best performing with p-value < 0.05. What's more, the improvement of the baseline methods varies across different datasets (e.g. referring to the underlined results), while our method offers a general denoising approach that achieves stable enhancements.

    The superiority of GNN-based models over classical models (BPRMF and Triple2vec) clearly demonstrates the significance of graph structure in modeling interactions and learning representations for within-basket recommendation tasks. However, we observed that the GNN-based method BasConv, which directly incorporates basket behavior, did not perform as well as expected. This observation suggests that poor-quality basket-item interactions can hinder representation learning and negatively impact performance.
    
    Next basket recommendation methods TIFU-KNN, as well as CLEA, exhibited strong performance on several metrics as TIFU-KNN reaches the second-best results with recall and hit rate on ValuedShopper confirming the importance of the usage of the shopping history in basket recommendation. TIFU-KNN's poor performance on NDCG indicates that non-neural network-based methods tend to overlook the order of recommendations to some extent. Additionally, we find that NBR methods heavily rely on the number of purchased baskets so that it boosts better performance on ValuedShopper dataset where each user has more baskets on average.
    
   It could be observed that methods such as SGL and CLEA which take denoising into account achieve better performance, showing the necessity of eliminating the effect of the noisy interaction in basket recommendation.
   Comparing SGL and CLEA, CLEA achieves better results since it focuses on denoising in the basket while SGL performs contrastive learning only on the user-item view, which proves the important role of basket denoising.

\subsection{Ablation Study (RQ2) }
In this section, we investigate the effectiveness of the proposed \model~ by evaluating the impact of different components. We denote the complementary basket hypergraph as B-I, the within-behavior contrastive learning with consistency-aware augmentation on the user-item interaction graph and basket view hypergraph as CA, and the cross-behavior contrastive learning as CL Fusion. We replace the consistency-aware augmentation with random edge perturbations and denote this model as $\model_{random}$. What's more, $\model_{add}$ represents the \model~ without hyperparameter $r$. Based on the typical user-item interaction graph, the checkmark under the corresponding module in Table \ref{tab:ablation} indicates whether this module was incorporated into the model. It can be seen that all the components are reasonably designed and essential to the final performance. We can observe that when any one of
these components is removed, the performance drops accordingly on all metrics. The findings could be summarized as follows:
\begin{itemize}[leftmargin=*]
    \item Incorporating the basket hypergraph leads to a noticeable enhancement in performance, confirming that basket behavior offers additional valuable information for representation learning.
    \item Note that when integrating user purchase behavior and basket behavior, the utilization of cross-behavior contrastive learning yields the most significant improvement in recommendation performance compared to directly combining the embeddings learned from both behaviors (e.g., achieving a 10.59\% increase in NDCG). This finding shows that cross-behavior contrastive learning contributes to reducing the effect of irrelevant information in the views of user purchase behavior and basket behavior by keeping the invariant and essential semantics of the items, which verifies the effectiveness of capturing the cooperative association between different views.
    \item The experimental results with random edge perturbations augmentation \model$_{random}$  and consistency-aware augmentation indicate that the consistency-aware augmentation will better identify the noise compared to a random structural augmentation and benefit the cross-behavior and within-behavior contrastive learning, which is consistent with our motivation in Section \ref{subsection:consistency aug}
    \item Compared with \model, the performance of \model$_{add}$ gets worse, which indicates that loosely adding the predictions of two separate views will include noise and ultimately results in suboptimal performance.

\end{itemize}

\begin{figure}[t]
     \centering
     \begin{subfigure}[t]{0.23\textwidth}
         \centering
         \includegraphics[width=1\textwidth]{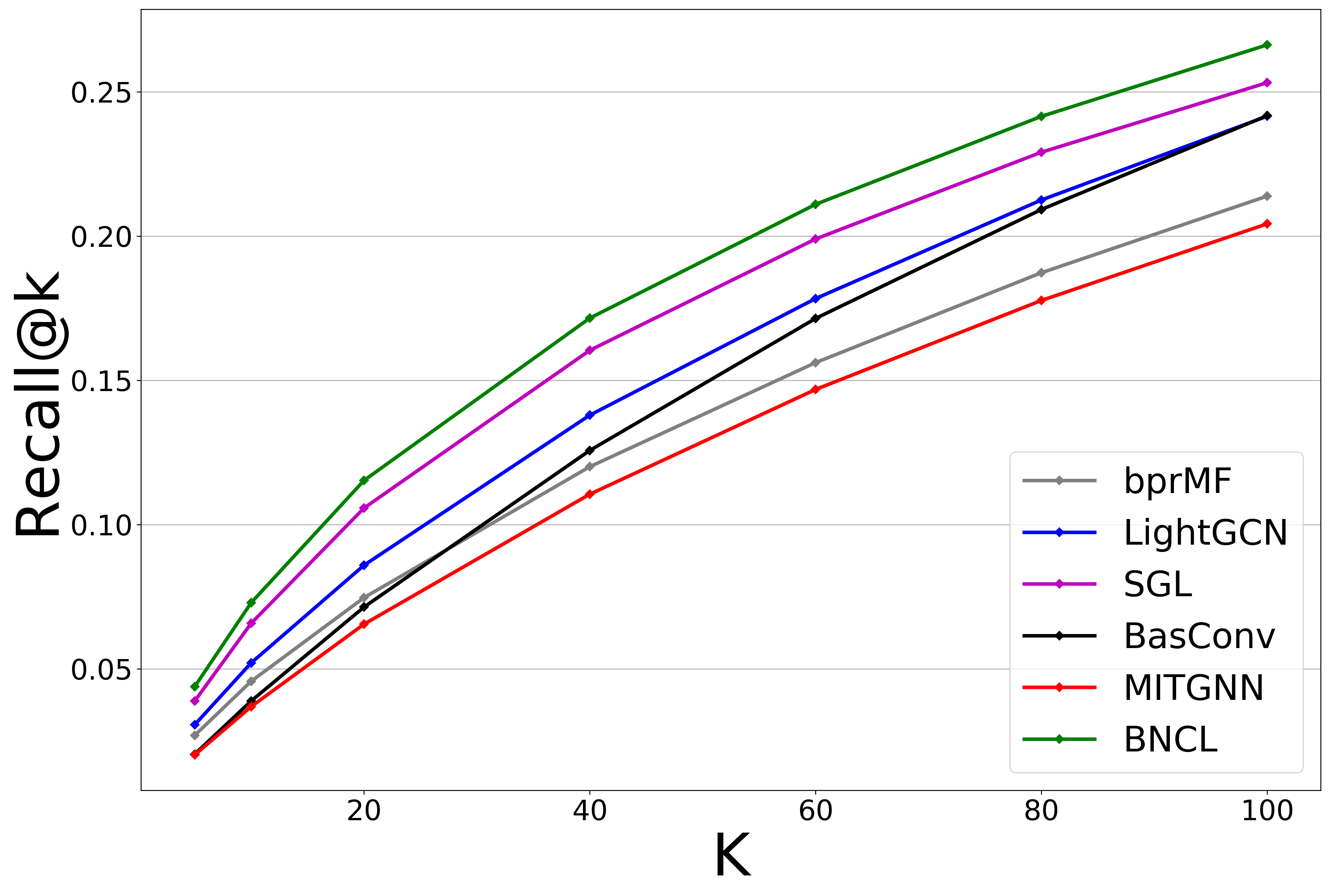}
         \caption{Recall@K}
         \label{fig:k_recall}
     \end{subfigure}
     \begin{subfigure}[t]{0.23\textwidth}
         \centering
         \includegraphics[width=1\textwidth]{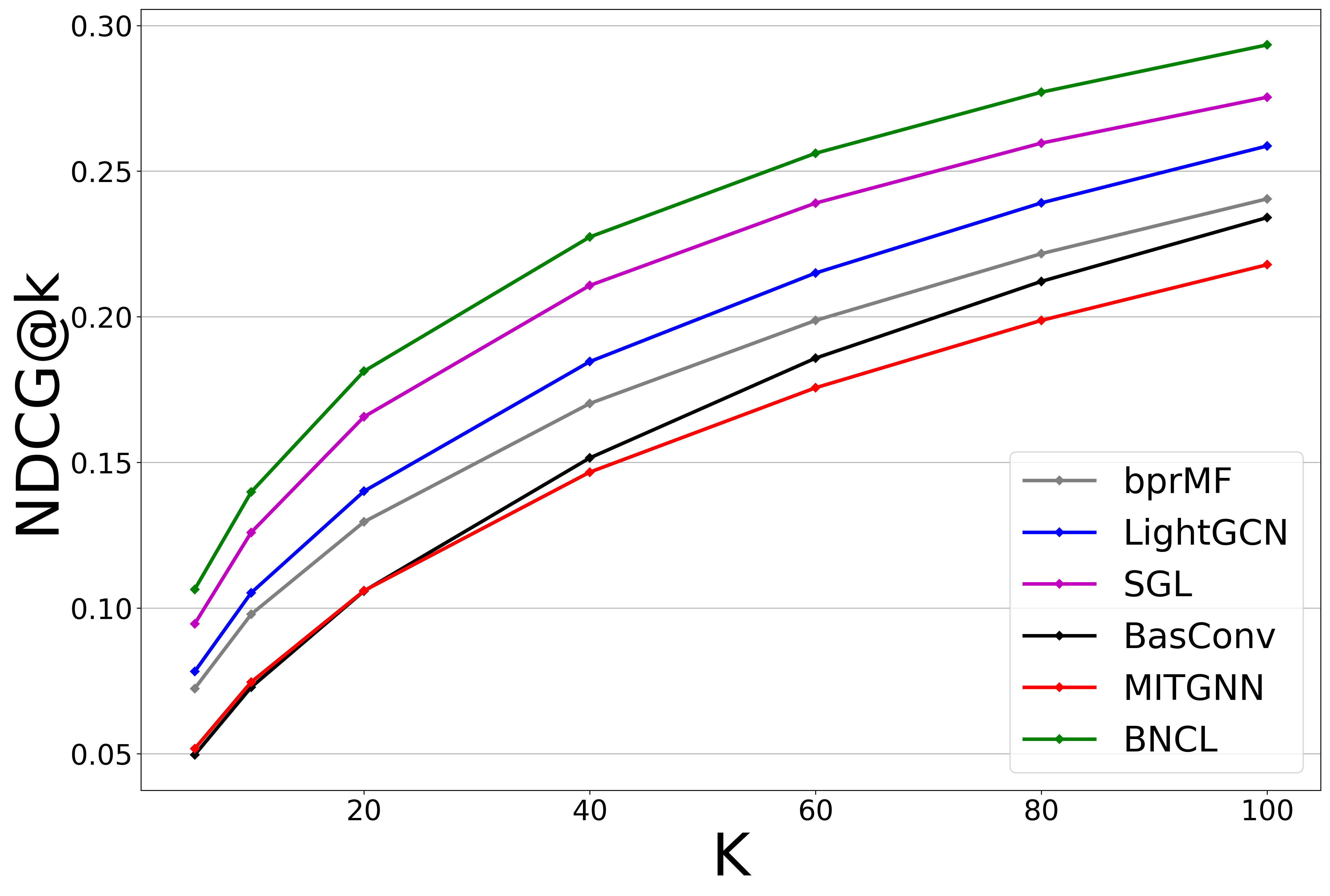}
         \caption{NDCG@K}
         \label{fig:k_ndcg}
     \end{subfigure}
        \caption{Performance on two metrics of \model~ and a few most representative baselines w.r.t. K in the range of $[5, 10, 20, 40, 60, 80, 100]$ on Instacart.} 
        \label{fig:dif_k}
\end{figure}
 \begin{table}[t]
\centering
\caption{Ablation Study of \model~on Instacart.}

\label{tab:ablation}
\scalebox{0.8}{
\begin{tabular}{ccc|cccc}
\hline
B-I &  CA  & CL Fusion    & Recall@60                & Precision@60            & HR@60           & NDCG@60             \\
\hline
- &  -             & -             & 17.83         & 2.33          & 70.31         & 21.50                   \\
$ \checkmark$ &  -             & -             & 18.28         & 2.39          & 70.77        & 21.62                 \\
$ \checkmark$ &   -           & $\checkmark$             & 20.17        & 2.64          & 73.50        & 23.91        \\

$ \checkmark$ & $\checkmark$              & -            & 20.87        & 2.73        &    73.509     &  25.22    \\ \hline
\multicolumn{3}{c|}{$\model_{random}$}              & 20.56          & 2.70      & 72.48         & 25.29        \\

\multicolumn{3}{c|}{$\model_{add}$}    & 20.02 & 2.67& 72.07 & 25.30\\

\multicolumn{3}{c|}{\model}    & \textbf{21.10} & \textbf{2.77} & \textbf{74.67} & \textbf{25.62}\\

\hline
\end{tabular}
}

\end{table}

\subsection{Case Study}
\subsubsection{Genralization Study (RQ3) }
We conduct experiments to verify the robustness and the generalization ability of the proposed \model. First, we test \model~with different $K$ in the range of [5, 10, 20, 40, 60, 80, 100]. The results are presented in Figure \ref{fig:dif_k}, which shows that on all metrics, with the increasing value of $K$, our model consistently outperforms the baseline methods.

Second, we test our model with the different message-passing backbones on the user-item interaction graph. In the typical recommendation algorithms, user-item interaction data is often modeled as a user-item bipartite graph like what we employed in the user-item interaction graph part. Numerous algorithms have focused on representation learning using user-item graphs \cite{berg2017graph, wei2019mmgcn}. Fism \cite{kabbur2013fism} proposed to improve the representation learning by training with the similarity matrix between items and for each training sample, it removes the direct link between the current user and its positive items when calculating the objective. We adopt Fism message-passing method and MF method on the user-item interaction modeling part of \model~instead of the LightGCN message-passing to learn the embedding. The results on Istacart dataset are shown in Table \ref{tab:backbone}, we can find that our model still performs the best even though the backbone has changed, which demonstrates the robustness and stability of our method.

\begin{table}[t]
\centering
\caption{Experimental comparisons of \model~on Instacart with different backbones.}

\label{tab:backbone}
\scalebox{0.85}{
\begin{tabular}{cccccc}
\hline
Backbone                                                                               & Recall@60               & Precision@60                              & HR@60               & NDCG@60                        \\ \hline

\text{ BPRMF}& 15.61           & 2.03                         & 66.05            & 19.88                  \\ 
\text { BPRMF-BNCL }                   &\textbf{16.72  }          & \textbf{2.51   }                        & \textbf{67.69}           & \textbf{20.39}              \\  \hline
\text { Fism }          & 15.73            & 2.06 & 64.33            & 20.92             \\ 
\text { Fism-BNCL }                                                                                    & \textbf{16.57} & \textbf{2.18}                       & \textbf{65.23}  & \textbf{22.65} \\ \hline
\text{ LightGCN}& 17.83 & 2.33& 70.31& 21.50           \\ 
\text { LightGCN-BNCL }                 & \textbf{21.10} & \textbf{2.77} & \textbf{74.67} & \textbf{25.62}  \\  \hline
\end{tabular}}
\end{table}

\subsubsection{Denoising Capability (RQ4) }
\begin{figure}[t]
     \begin{subfigure}[t]{0.23\textwidth}
         \centering
         \includegraphics[width=1\textwidth]{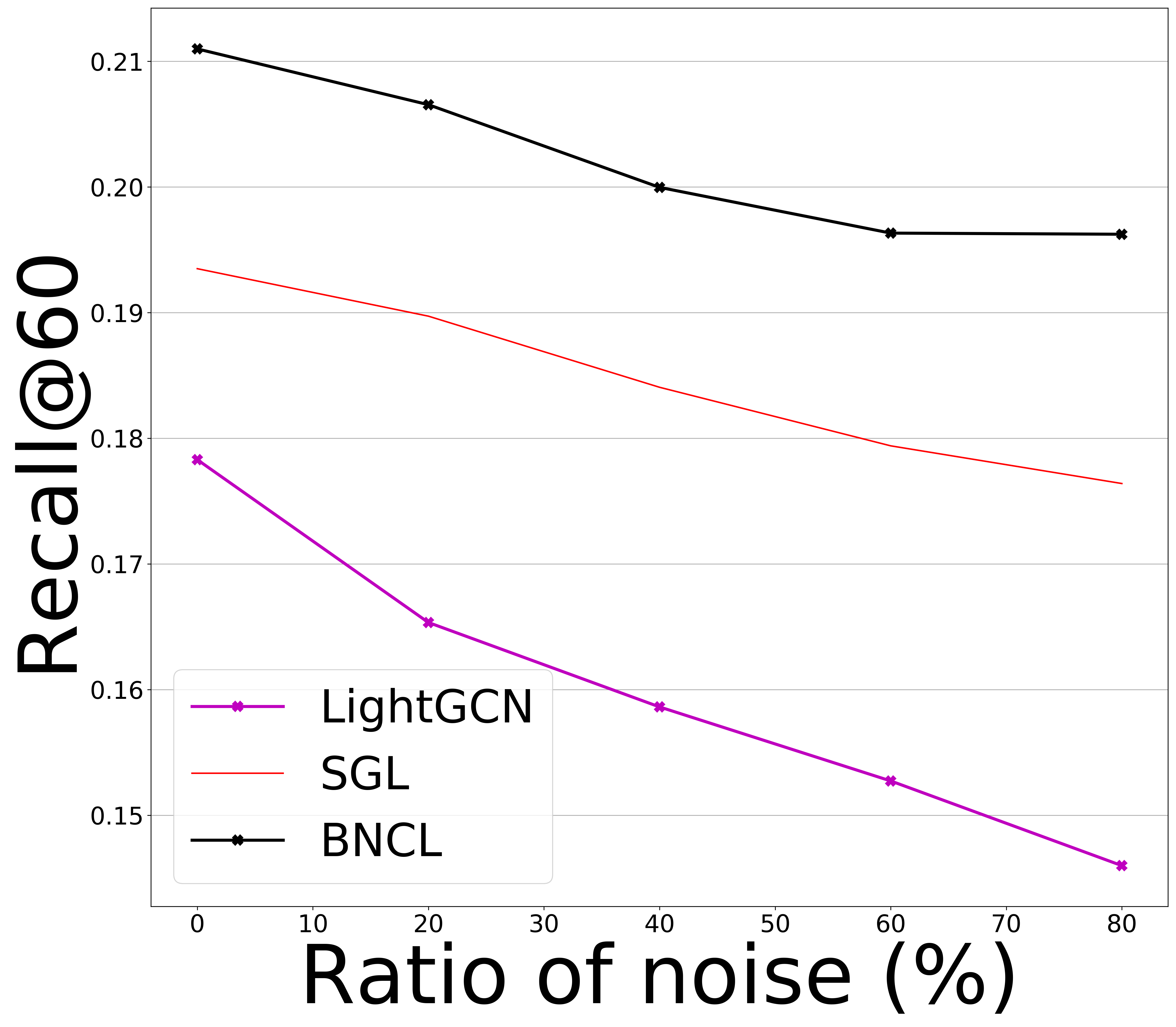}
         \caption{Recall@60}
         \label{fig:erm_domain}
     \end{subfigure}
     \hfill
     \begin{subfigure}[t]{0.23\textwidth}
         \centering
         \includegraphics[width=1\textwidth]{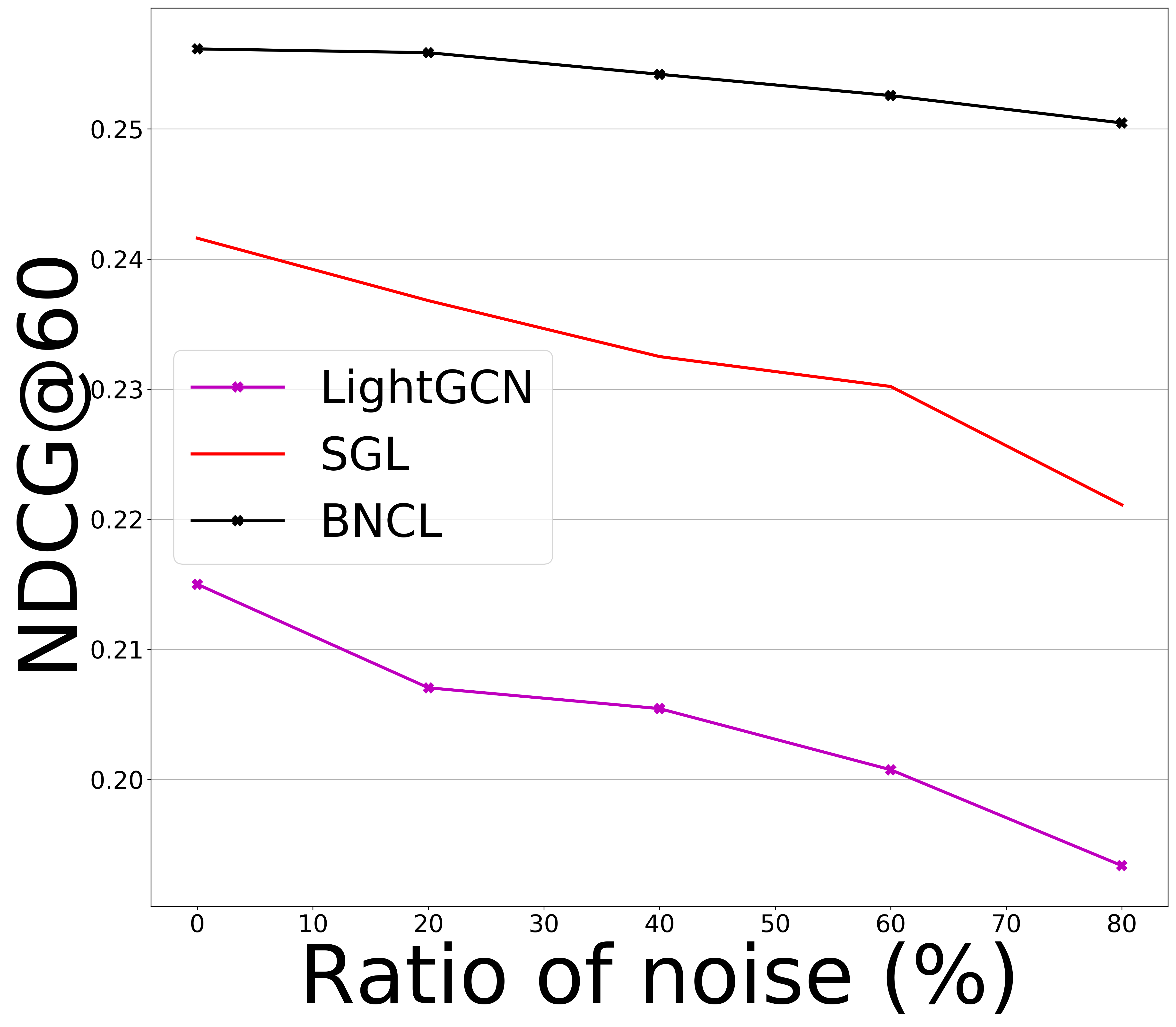}
         \caption{NDCG@60}
         \label{fig:scl_domain}
     \end{subfigure}\vspace{-3mm}
        \caption{Performance on two metrics of {\model} and LightGCN with varying ratios of noise added on Instacart.} 
        \label{fig:noise}

\end{figure}

In this section, we further investigate the denoising performance of the proposed method. We introduce varying levels of noise to the user-item interaction graph and the basket hypergraph, specifically adding 20\%, 40\%, 60\%, and 80\% noisy interactions, and then observe the performance of \model~. we consistently observe that \model~outperforms both LightGCN and SGL across all noise ratios. As depicted in Figure \ref{fig:noise}, we note that as the ratio of added noise increases from 20\% to 80\%, our method experiences a mere 6.99\% decline in Recall@60, whereas LightGCN's performance drops by 18.12\% and SGL's by 9.00\%. Similarly, the drop in NDCG@60 for our method is only 2.22\%, whereas LightGCN experiences a 10.06\% decline and SGL an 8.49\% decline. These results serve as further evidence of the robustness of our model in handling noisy interactions. The evaluation under more advanced attack algorithms \cite{huai2020malicious,fang2018poisoning} is left as future work.

\begin{spacing}{0.99}
\section{Related Work}
In this section, we briefly review the related work on basket recommendation and contrastive learning for recommendation.
\vspace{-3mm}
\label{section:related_work}
\subsection{Basket Recommendation}
Basket recommendation (BR) \cite{gatzioura2014case,le2017basket} is to recommend a set of items that are mostly possible purchased by targeted users based on their shopping records. The basic idea is to capture correlations and perform the prediction through
Collaborative Filtering (CF) methods \cite{he2017neural,wei2020fast,zhang2021causal}
 or Markov Chain (MC) methods \cite{rendle2010factorizing,wang2015learning}. FPMC \cite{rendle2010factorizing} is proposed to capture both sequential effects and long-term user taste where each user-specific transition is modeled by an underlying MC. 
Focus on predicting the items for the user’s next baskets, DREAM \cite{yu2016dynamic} used an LSTM network \cite{hochreiter1997long} to capture the series features of the basket sequence. 
Some studies explore within-basket recommendations that also consider the content of the current basket.
Triple2vec \cite{wan2018representing} improves the within-basket recommendation by constructing the training samples as triples of the user and the items. DBFM  \cite{LI2022116383} contributes a basket recommendation solution based on factorization with a deep neural network. 
PerNIR \cite{ariannezhad2023personalized} models the short-term interests of users represented by the current basket, as well as their long-term interests to address the task.
Recently, GNN has shown its great potential to capture the interactions among the user and the item in representation learning \cite{scarselli2008graph,berg2017graph,wang2019neural}. 
Basconv \cite{liu2020basconv} is proposed to capture heterogeneous interaction signals on a UBI graph by designing three different aggregators for user, basket and item entities. MITGNN \cite{liu2020basket} combines the translation-based model
with the GNN to improve within-basket
recommendation via retrieving the multi-intent pattern. 
In practice, noisy interactions are easy to occur during shopping behaviors and will hinder the capture of users' preferences and item properties while few works explicitly consider denoising in BR. For NBR, CLEA \cite{qin2021world} used a denoising generator to denoise the baskets and then extract relevant items to enhance recommendation performance. However, the identification of noisy interactions and the elimination of their impact on heterogeneous behaviors in basket recommendation have been overlooked in the current literature.

\vspace{-3mm}
\subsection{Contrastive Learning for Recommendation}
Contrastive learning aims to learn representation by minimizing the distance of positive instances while making negative instances far apart in the representation space \cite{chen2020simple}, which has achieved great success on graph \cite{you2020graph,hassani2020contrastive,hangraph} and hypergraph \cite{wei2022augmentations} representation learning, as well as the application in recommender systems \cite{wu2021self} and dense retrieval \cite{Karpukhin2020DensePR,Zeng2022CurriculumLF,Zeng2023APD}. A Contrastive multi-view graph representation
learning algorithm \cite{hassani2020contrastive} is introduced for learning both node and graph-level representations by contrasting structural views of graphs.
CLRec \cite{zhou2021contrastive} bridged the theoretical gap between contrastive learning objective and traditional recommendation objective, which showed that directly performing contrastive learning can help to reduce the exposure bias. Neighborhood-enriched Contrastive Learning (NCL) \cite{lin2022improving} explicitly incorporated the potential neighbors into contrastive pairs by introducing the neighbors of a user (or an item) from graph structure and semantic space respectively. CMP-PSP \cite{wu2022multi} effectively leveraged contrastive multi-view learning and pseudo-siamese networks to mitigate data sparsity and noisy interactions.
CCFCRec \cite{zhou2023contrastive} adopts contrastive collaborative filtering for cold-start item recommendation which applies contrastive learning to transfer the co-occurrence signals to the content CF module. KACL \cite{wang2023knowledge} performs contrastive learning across the user-item interaction view and KG view to include the knowledge graph in the recommendation while eliminating the noise it may introduce.
Despite these advancements, the potential of contrastive learning in denoising within basket recommendation remains underexplored. Our work aims to harness the power of contrastive learning in behavior denoising and integrating diverse behaviors.
 
\vspace{-3mm}
\section{Conclusion}
\label{section:conclusion}
In this paper, we formulate the basket recommendation by integrating the basket behaviors into user-item interaction modeling with a light hypergraph message passing schema and a joint self-supervised learning paradigm. We systematically illustrate the noise issues in the basket recommendation problem. A general basket recommendation framework \model~via noise-tolerated contrastive learning is proposed accordingly to improve robustness against the noise in BR. To be specific, we suppress the cross-behavior noise by making use of additional supervision signals with cross-behavior contrastive learning. Then to inhibit the within-behavior noise in the user and basket interactions, we propose to exploit invariant properties of the recommenders w.r.t augmentations through within-behavior contrastive learning. In addition, a novel consistency-aware augmentation approach is designed to better identify noisy interactions by comprehensively considering the two types of interactions. Extensive experimental results over three datasets on within-basket recommendation task show that our proposed method outperforms state-of-the-art baselines in terms of various ranking metrics.

One direct extension of our work is using \model~as an effective tool to suppress the noise in multi-view learning and help the model fusion.  Additionally, incorporating temporal dynamics and the order of the baskets into our current framework is a promising avenue for future research. Moreover, we'd like to test our method with more advanced backbones and self-supervised learning strategies. We're also interested in evaluating the bias \cite{wei2021model} and fairness \cite{wei2022comprehensive} of our approach under various settings.

\vspace{-3mm}
\begin{acks}
This work is supported by National Science Foundation under Award No. IIS-1947203, IIS-2002540, IIS-2117902, IIS-2137468, Agriculture and Food Research Initiative (AFRI) grant no. 2020-67021-32799/project accession no.1024178 from the USDA National Institute of Food and Agriculture, and IBM-Illinois Discovery Accelerator
Institute - a new model of an academic-industry partnership designed to increase access to technology education
and skill development to spur breakthroughs in emerging areas of technology. The views and conclusions are those of the authors and should not be interpreted as representing the official policies of the funding agencies or the government.
\end{acks}

\end{spacing}
\bibliographystyle{ACM-Reference-Format}
\bibliography{output}

\end{document}